\documentclass[english,aps, prl, twocolumn, superscriptaddress]{revtex4}

\usepackage[T1]{fontenc}
\usepackage[latin9]{inputenc}
\usepackage{graphicx}
\usepackage{amssymb}

\makeatletter

\usepackage{babel}
\makeatother

\begin{document}

\title{Cavity grid for scalable quantum computation with superconducting
circuits}

\author{Ferdinand Helmer }

\email{Ferdinand.Helmer@physik.lmu.de}

\affiliation{{\small Department of Physics, CeNS, and ASC, Ludwig-Maximilians-Universität,
Theresienstrasse 37, D-80333 Munich, Germany}}

\author{Matteo Mariantoni}

\affiliation{{\small Walther-Meißner-Institut, Bayerische Akademie der Wissenschaften,
Walther-Meißner-Strasse 8, D-85748 Garching, Germany}}
\affiliation{{\small Department of Physics, Technische Universit\"{a}t M\"{u}nchen - James-Franck-Str., 85748 Garching, Germany}}

\author{Austin G. Fowler}

\affiliation{{\small Institute for Quantum Computing, University of Waterloo,
Waterloo, ON, Canada}}

\author{Jan von Delft}

\affiliation{{\small Department of Physics, CeNS, and ASC, Ludwig-Maximilians-Universität,
Theresienstrasse 37, D-80333 Munich, Germany}}

\author{Enrique Solano}

\affiliation{{\small Department of Physics, CeNS, and ASC, Ludwig-Maximilians-Universität,
Theresienstrasse 37, D-80333 Munich, Germany}}

\affiliation{{\small Secci\'{o}n F\'{\i}sica, Departamento de Ciencias, Pontificia Universidad Cat\'{o}lica del Per\'{u}, Apartado Postal 1761, Lima, Peru}}

\author{Florian Marquardt}

\affiliation{{\small Department of Physics, CeNS, and ASC, Ludwig-Maximilians-Universität,
Theresienstrasse 37, D-80333 Munich, Germany}}

\begin{abstract}
\noindent We propose an architecture for quantum computing based on
superconducting circuits, where on-chip planar microwave resonators
are arranged in a two-dimensional grid with a qubit at each intersection.
This allows any two qubits on the grid to be coupled at a swapping
overhead independent of their distance. We demonstrate that this approach
encompasses the fundamental elements of a scalable fault-tolerant
quantum computing architecture.
\end{abstract}
\maketitle
\newcommand{\ket}[1]{\left|#1\right\rangle }

\newcommand{\bra}[1]{\left\langle #1\right|}

\newcommand{\s}{\hat{\sigma}}

\section{Introduction}
Superconducting circuits are promising candidates for scalable quantum information
processing \cite{2001_04_MakhlinShnirmanSchoen_RMP,1998_01_Devoret_CooperPairBox,1999_04_Nakamura_CooperPairBox,2000_10_Mooij_FluxQubit_Experiment,2002_05_Vion_Quantronium,2002_08_Martinis_PhaseQubit,2005_03_Majer_TwoQubitsPRL,2006_09_Martinis_EntanglingPhaseQubits,2006_12_Clarke_TunableCoupling,2007_05_Nakamura_SwitchableCoupling}.
This route was further strengthened with the advent of circuit quantum electrodynamics. Starting with 
early proposals for implementing the quantum-optical Jaynes-Cummings model in the context of superconducting circuits \cite{2001_Marquardt_CooperBox,2001_BuissonHekking_Cat,2003_YouNori_CircuitQED}, this research direction became a major topic after it was pointed out that on-chip microwave transmission line resonators could be coupled to superconducting qubits \cite{2004_02_BlaisEtAl_CavityProposal}. Since then, a series of ground-breaking experiments have demonstrated these concepts \cite{2004_09_WallraffEtAl_MicrowaveCavity,2004_09_MooijFluxQubitJC,2006_03_Semba_JapaneseCircuitQED}, including achievements like dispersive qubit readout \cite{2005_08_Wallraff_PRL_UnitVisibility}, photon number splitting \cite{2007_02_Yale_PhotonNumberSplitting}, single-photon generation \cite{2007_Yale_nature_Single_Photon_Source}, or lasing by a single artifical atom \cite{2007_10_Nakamura_CPB_Laser}.

Recent experiments \cite{2007_09_Sillanpaa_TwoQubits,2007_09_Majer_TwoQubits}
have advanced to coupling two qubits via the cavity, yielding a flip-flop
(XY) interaction permitting two-qubit gates. If multiple qubits share
one cavity, arbitrary qubit pairs could be selectively coupled \cite{2006_08_Wallquist_TunableCavity},
which outperforms nearest-neighbor setups (no swapping overhead or
disruption by single unusable qubits). However, moving towards more
qubits requires suitable novel architectures.

\begin{figure}
\includegraphics[width=\columnwidth]{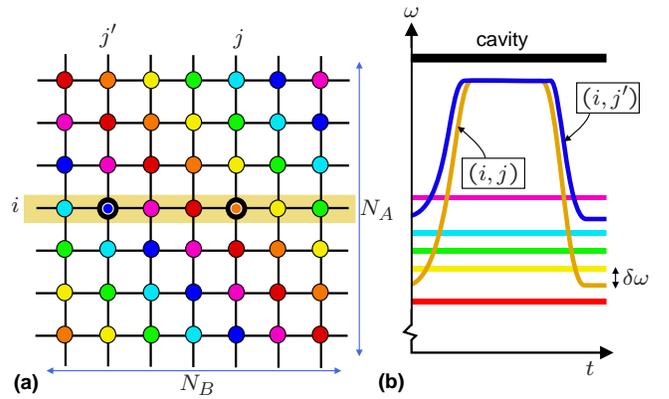}

\caption{\textbf{\label{SimpleGrid}}Schematic cavity grid setup. (a) The 2D
cavity grid, with qubits depicted as circles and cavities shown as
lines. Qubit $(i,j)$ sits at the intersection of cavities $i$ and
$j$. Colors distinguish the transition frequencies, which differ
within any column or row (in the `idle state'). (b) A two-qubit operation
is induced by tuning two qubits into mutual resonance to exploit the
cavity-assisted dispersive coupling. }

\end{figure}

In this Letter, we present and analyze an architecture that builds
on these elements and extends them into the second dimension, by forming
a crossbar-like geometry of orthogonal microwave resonators, with
qubits sitting at the intersections (Fig.~\ref{SimpleGrid}). The
global coupling within each row and column makes this setup distinct
from existing proposals for array-like arrangements in ion traps \cite{2000_04_CiracZoller_IonArray,2002_Wineland_QuantumCCD},
optical lattices \cite{1999_03_Jaksch_OpticalLatticesqC}, semiconductor
spins \cite{2005_12_ZollerMarcus_QDotArray} or superconductors \cite{2001_04_MakhlinShnirmanSchoen_RMP,2005_Wallquist_QubitNetwork}. We show (i) how to
couple any two qubits on the grid, (ii) with minimal swapping overhead
using (iii) an appropriately chosen ('Sudoku'-style) frequency distribution, and (iv) suggest a
scalable fault-tolerant quantum computing architecture. 

Before we turn to a description of our proposal, we note that experiments right now are obviously still struggling to improve
the fidelity of single- and two-qubit operations for superconducting
qubits, and this painstaking work is crucial for further progress
in the whole field. Nevertheless, the effort going into this endaveour
is ultimately justified by the long-term goal of implementing large-scale
circuits able to perform nontrivial quantum computation tasks, where
the numbers of qubits may run into the thousands. While present-day
experiments are still very far removed from this goal, it is
worthwile to develop architectures that couple more than a handful
of qubits in a nontrivial setup, and which represent a challenging
medium term goal for the experiments to strive for. We will demonstrate
that parameters (dephasing times, coupling strengths etc.) near those
that are available nowadays would allow for a first proof-of-principle
experiment in our proposed architecture, and further progress in the
perfection of single qubits will enable truly useful larger scale
versions. The basic ideas behind our scheme are sufficiently general
so as to permit replacing individual building blocks (particular qubit
types, two-qubit gates etc.) by improved versions that might be developed
within the coming years.

In addition, we would like to emphasize that even though any working
set of universal one- and two-qubit operations permits to implement
arbitrary computations in principle, it is by no means clear that
the resulting generic implementation is efficient. Rather, in order to make
the most efficient use of resources, it is mandatory to come up with
larger scale schemes that exploit the particular features of a given
physical realization. In this sense, our proposal is similar in spirit
to previous proposals for other physical systems that envisaged how
well-known elementary operations could be extended to an efficient
two-dimensional architecture \cite{2000_04_CiracZoller_IonArray,2002_Wineland_QuantumCCD,1999_03_Jaksch_OpticalLatticesqC}.

\section{Basic Architecture}
The cavity grid consists of
cavity modes belonging to $N_{A}$ horizontal (A) and $N_{B}$ vertical
(B) cavities, $\hat{H}_{{\rm cav}}=\sum_{j=1}^{N_{A}}\hbar\omega_{j}^{A}\hat{a}_{j}^{\dagger}\hat{a}_{j}+\sum_{j=1}^{N_{B}}\hbar\omega_{j}^{B}\hat{b}_{j}^{\dagger}\hat{b}_{j}$,
coupled to one qubit of frequency $\epsilon_{ij}$ at each intersection
$(i,j)$, generalizing \cite{2004_02_BlaisEtAl_CavityProposal} to
a 2D architecture:

\begin{eqnarray}
\hat{H}_{{\rm cav}-{\rm qb}} & = & \sum_{i,j}\hat{n}_{ij}[g_{ij}^{A}(\hat{a}_{i}+\hat{a}_{i}^{\dagger})+g_{ij}^{B}(\hat{b}_{j}+\hat{b}_{j}^{\dagger})].\label{Hamiltonian}\end{eqnarray}
For definiteness we consider charge (or transmon) qubits, unless noted
otherwise. Then the couplings $g_{ij}^{A(B)}$ between the horizontal
(vertical) cavity mode $i\,(j)$ and the dipole operator $\hat{n}_{ij}$
of qubit $(i,j)$ depend on the detailed electric field distribution
and geometry of the qubit. Eq.~(\ref{Hamiltonian}) leads to the
Jaynes-Cummings model and the cavity-mediated interaction between
qubits \cite{2004_02_BlaisEtAl_CavityProposal}. It can be realized
in different ways: A capacitive coupling was demonstrated for charge
\cite{2004_09_WallraffEtAl_MicrowaveCavity} (or 'transmon' \cite{2007_02_Yale_PhotonNumberSplitting,2007_09_Majer_TwoQubits})
and phase qubits \cite{2007_09_Sillanpaa_TwoQubits}, while for flux
qubits \cite{2004_09_MooijFluxQubitJC} $\hat{n}_{ij}$ is the magnetic
moment coupling to the magnetic field.

It is well-known \cite{2004_02_BlaisEtAl_CavityProposal,2007_09_Majer_TwoQubits}
that the Hamiltonian (\ref{Hamiltonian}) induces an effective flip-flop
interaction of strength $J_{\alpha\beta}=g_{\alpha}g_{\beta}(\Delta_{\alpha}+\Delta_{\beta})/(2\Delta_{\alpha}\Delta_{\beta})$
between each pair of qubits $(\alpha,\beta)$ in the same cavity (for
couplings $g_{\alpha(\beta)}$ and detunings from the cavity $\Delta_{\alpha(\beta)}$,
in the dispersive limit $\left|g\right|\ll\left|\Delta\right|$):

\begin{equation}
\hat{H}_{\alpha\beta}^{{\rm flip-flop}}=J_{\alpha\beta}\left(\hat{\sigma}_{\alpha}^{+}\hat{\sigma}_{\beta}^{-}+{\rm h.c.}\right).\label{Hflipflop}\end{equation}
In the computational `idle state' these interactions have to be effectively
turned off by detuning all the qubits from each other. This requires
a detuning $\delta\omega\gg J$ to avoid spurious two-qubit operations.
Thus, the number $N$ of qubits in a linear array is strongly restricted
\cite{2006_08_Wallquist_TunableCavity}, since a frequency interval
of order $N\delta\omega$ is required. In the present 2D architecture,
this constraint is considerably relaxed. The required frequency range
is reduced from $N\delta\omega$ to $\sqrt{N}\delta\omega$ (where
$N$ is the total number of qubits), while still ensuring a spacing
of $\delta\omega$ within each cavity (the constraints being similar
to the rules of the game {}``Sudoku''). This allows for grids with
more than $20\times20=400=N$ qubits, for realistic parameters. Figure
\ref{SimpleGrid} shows an acceptable frequency distribution. An extension
to a fully scalable setup is discussed at the end of this paper.

\section{One-qubit operations}
We briefly review some ingredients
that have already been implemented \cite{2004_09_WallraffEtAl_MicrowaveCavity,2004_09_MooijFluxQubitJC,2005_08_Wallraff_PRL_UnitVisibility,2006_03_Semba_JapaneseCircuitQED,2007_02_Yale_PhotonNumberSplitting}.
Operations on a selected qubit can be performed via Rabi oscillations
\cite{2005_08_Wallraff_PRL_UnitVisibility} using a microwave pulse
resonant with the qubit at $\epsilon_{ij}/\hbar$ but detuned from
the cavity and all other qubits in the same cavity. Rotations around
the $z$-axis can be performed via AC Stark shift \cite{2007_09_Majer_TwoQubits},
or by tuning the qubit frequency temporarily (see below). The cavities
can be used for fast dispersive QND readout \cite{2005_08_Wallraff_PRL_UnitVisibility}
of single qubits tuned close to the readout frequency or multiplexed
readout of several qubits at once \cite{2004_02_BlaisEtAl_CavityProposal,2007_09_Majer_TwoQubits}.
\begin{figure}
\includegraphics[width=\columnwidth]{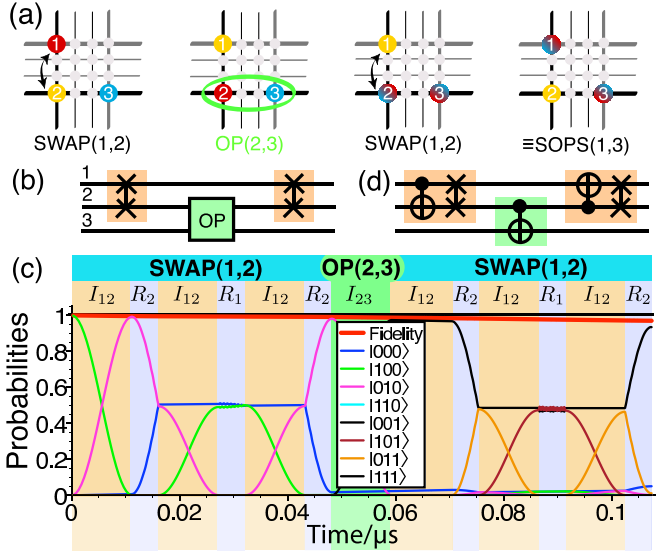}

\caption{\label{simulation}Operations between arbitrary qubits on the grid
($I_{ij}$ denotes an iSWAP gate between qubits $i$ and $j$, and
$R_{i}$ an $x$ rotation by $-\pi/2$). (a) Sequence of operations
for a two-qubit gate between two qubits ($1$ and $3$), via an auxiliary
qubit ($2$). (b) Corresponding quantum circuit, where each $\mbox{SWAP}$
has to be decomposed into three $\mbox{iSWAP}$s and local gates,
Eq. (\ref{eq:SWAPit}). (c) Master equation simulation of the full
evolution for an operation according to (b), including relaxation
and dephasing. The evolution of all three-qubit probabilities is shown
together with the fidelity (topmost curve), for presently available
experimental parameters. (d) For the important case $\mbox{OP}=\mbox{CNOT}$,
a speed-up can be obtained by noting that each $\mbox{SWAP}/\mbox{CNOT}$
pair can be implemented using a single $\mbox{iSWAP}$ and local gates
(see \cite{2003_03_SchuchSiewert_XYgates}). }

\end{figure}

\section{Tunability}
\begin{figure}
\begin{centering}
\includegraphics[width=\columnwidth]{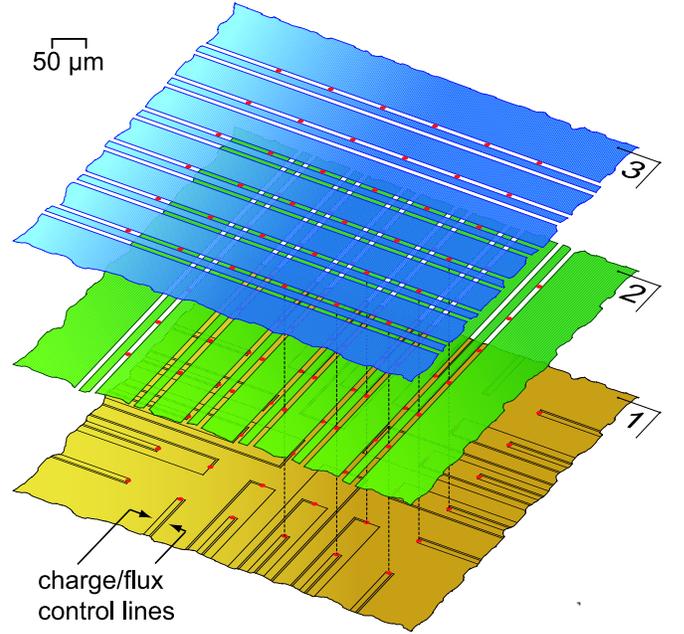}
\par\end{centering}

\caption{\label{multilayersetup}A possible multilayer architecture.\textbf{
}The layers 2 and 3 with coplanar wave guides are positioned above
a `control line layer' 1. The qubit positions are indicated as red
dots within each layer only for reference (they would be fabricated
above layer 3). }

\end{figure}

Additional charge and flux control
lines (Fig.~\ref{multilayersetup}) reaching each qubit are needed
for tunability. For split-junction charge qubits \cite{1999_04_Nakamura_CooperPairBox,2001_04_MakhlinShnirmanSchoen_RMP},
locally changing the magnetic flux sweeps the energy splitting $\epsilon_{ij}=E_{{\rm J}}(\Phi_{ij})$
(see \cite{2006_02_JenaGroup_FourQubits,2007_09_Sillanpaa_TwoQubits}),
while keeping the qubit at the charge degeneracy point (to which it
has been tuned via a separate charge gate line).\textbf{ }This ensures
maximum coherence through weak coupling to $1/f$ noise, although
this requirement is relaxed in the new {}``transmon'' design \cite{2008_03_Houck_TransmonRelaxation}.
Individual addressability introduces some hardware overhead, but is
essential both for two-qubit gates and for compensating fabrication
spread. %

\section{Two-qubit gates}
We use the effective flip-flop
interaction of Eq.~(\ref{Hflipflop}) (see \cite{1999_11_Imamoglu_QDotsCavityQED,2001_07_OsnaghiHaroche_DispersiveCoupling,2004_02_BlaisEtAl_CavityProposal,2007_Blais_PRA_QIP_with_circuit_QED})
to induce two-qubit gates. In the `idle state', the interaction is
ineffective, since the qubits are out of resonance, $\left|\epsilon_{\alpha}-\epsilon_{\beta}\right|\gg J$.
During the gate, the two qubit frequencies are tuned into mutual resonance
near the cavity frequency to increase $J$, see Fig.~\ref{SimpleGrid}~(b).
After a waiting time $t=\hbar\pi/(2|J|)$, this realizes the universal
two-qubit $\mbox{iSWAP}$ gate (demonstrated experimentally in \cite{2007_09_Majer_TwoQubits}),
which can be used to construct $\mbox{CNOT}$ and $\mbox{SWAP}$.
Each $\mbox{SWAP}(\alpha,\beta)$ operation in the protocol (Fig.
\ref{simulation}) can be decomposed into three $\mbox{iSWAP}$ gates
between qubits $\alpha$ and $\beta$ \cite{2003_03_SchuchSiewert_XYgates}:

\begin{equation}
\mbox{SWAP}=\mbox{iSWAP}\cdot\mbox{R}_{\beta}\cdot\mbox{iSWAP}\cdot\mbox{R}_{\alpha}\cdot\mbox{iSWAP}\cdot\mbox{R}_{\beta}.\label{eq:SWAPit}\end{equation}
Here $\mbox{R}_{\alpha}$ rotates qubit $\alpha$ by an angle $-\pi/2$
around the $x$-axis via a Rabi pulse.\textbf{ }Arbitrary gates between
any two qubits (e.g., $1$ and $3$) in different cavities can be
implemented via an intermediate qubit $2$ at the junction of two
orthogonal cavities containing $1$ and $3$ (see Fig. \ref{simulation}).
The sequence 

\begin{equation}
\mbox{SOPS}(1,3)\equiv\mbox{SWAP}(1,2)\,\mbox{OP}(2,3)\,\mbox{SWAP}(1,2),\label{SOPS}\end{equation}
leaves qubit $2$ unchanged and performs the desired operation {}``$\mbox{OP}$''
between $1$ and $3$. 

We simulated such an operation (Fig.~\ref{simulation}) for realistic
parameters. Relaxation and pure dephasing for each qubit $\alpha$,
with rates $\gamma$ and $\gamma^{\varphi}$, are modeled by a Lindblad
master equation (where $\hat{P}_{\alpha}=\ket{e_{\alpha}}\bra{e_{\alpha}}$
projects onto the excited state of qubit $\alpha$):

\begin{eqnarray}
\dot{\hat{\rho}} & = & -\frac{i}{\hbar}[\hat{H},\hat{\rho}]+\sum_{\alpha}(\mathcal{L}_{\alpha}^{\varphi}+\mathcal{L}_{\alpha}^{{\rm rel}})\hat{\rho}\,,\\
\mathcal{L}_{\alpha}^{\varphi}\hat{\rho} & = & \gamma^{\varphi}\left[2\hat{P}_{\alpha}\hat{\rho}\hat{P}_{\alpha}-\hat{P}_{\alpha}\hat{\rho}-\hat{\rho}\hat{P}_{\alpha}\right]\,,\\
\mathcal{L}_{\alpha}^{{\rm rel}}\hat{\rho} & = & \gamma\left[\s_{\alpha}^{-}\hat{\rho}\s_{\alpha}^{+}-\frac{1}{2}\s_{\alpha}^{+}\s_{\alpha}^{-}\hat{\rho}-\frac{1}{2}\hat{\rho}\s_{\alpha}^{+}\s_{\alpha}^{-}\right].\end{eqnarray}
We consider three qubits, where $(1,2)$ and $(2,3)$ are coupled
via flip-flop terms {[}see Eq.~(\ref{Hflipflop})] after adiabatic
elimination of the cavities. During two-qubit gates, the qubit energy
is ramped and will cross other qubit energies (Fig.~\ref{SimpleGrid}),
potentially leading to spurious population transfer to other qubits
if the process is too slow, while ramping too fast would excite higher
qubit levels. For a $10\,{\rm ns}$ switching time (during which a
sweep over $\delta\epsilon/\hbar=2\pi\cdot10\,{\rm GHz}$ is accomplished),
the probability of erroneous transfer during one crossing is estimated
to be less than $10^{-2}$ from the Landau-Zener tunneling formula,
and thus could be safely disregarded for the present simulation, where
energies were instead switched instantaneously. Although several crossings
may occur during one sweep, the scalable setup to be introduced further
below keeps this kind of error under control by having only eight
qubits per cavity\textbf{.} For the simulation we used the following
parameters: Initially, the qubit transition frequencies are at $\epsilon/\hbar=2\pi\cdot4,\,5,\,6\,{\rm GHz}$.
A resonant classical drive yields a Rabi frequency of $\Omega_{{\rm R}}=150\,{\rm MHz}$.
A qubit-cavity coupling $g=2\pi\cdot150\,{\rm MHz}$ and a detuning
$\Delta=2\pi\cdot1\,{\rm GHz}$ (from a cavity at $2\pi\cdot15\,{\rm GHz}$)
produce $J/\hbar=2\pi\cdot21\,{\rm MHz}$. The dephasing and decay
rates are $\gamma^{\varphi}=0.16\,{\rm MHz}$ and $\gamma=0.6\,{\rm MHz}$
(i.e. $T_{1}=1.7\,\mu s$ and $T_{\varphi}=6.3\mu s$), consistent
with recent experiments on transmon qubits \cite{2008_03_Houck_TransmonRelaxation}.
Note that in the idle state the actual $J$ is reduced by a factor
of $10$ (due to larger detuning from the cavity). Employing a qubit
spacing of $\delta\omega\sim500\,{\rm MHz}$, this yields a residual
coupling strength of $J^{2}/\hbar^{2}\delta\omega\sim0.4\,{\rm MHz}$,
which may be reduced further by refocusing techniques. To check the
accuracy of adiabatic elimination, we performed an additional simulation
of an $\mbox{iSWAP}$ operation between two qubits taking the cavity
fully into account, observing an error below the level brought about
by dissipation. 

A measure of the fidelity of the operation is obtained \cite{2004_Nielsen_Fidelities}
by computing $F(\hat{\rho}_{{\rm real}}(t),\hat{\rho}_{{\rm ideal}}(t))$,
where $F(\hat{\rho}_{1},\hat{\rho}_{2})\equiv{\rm tr}(\sqrt{\sqrt{\hat{\rho}_{1}}\hat{\rho}_{2}\sqrt{\hat{\rho}_{1}}})^{2}$,
and $\hat{\rho}_{{\rm ideal}}$ denotes the time-evolution in the
absence of dissipation. Fig.~\ref{simulation} shows a fidelity of
about $95\%$, confirming that presently achievable parameters suffice
for a first proof-of-principle experiment. 

We emphasize that the swapping overhead does not grow with the distance
between the qubits. Furthermore, multiple operations may run in parallel,
even if they involve the same cavities, provided no qubit is affected
simultaneously by two of the operations and the qubit pairs are tuned
to different frequencies. 

Here we have chosen the dispersive two-qubit gate that relies on proven
achievements. Faster resonant gates (e.g. CPHASE \cite{1999_12_Rauschenbeutel_CPHASE,2006_08_Wallquist_TunableCavity})
might be implemented, with a time scale on the order of $1/g$ instead
of $\Delta/g^{2}$. 

\section{Hardware}
For illustration, we discuss one out
of many conceivable setups (Fig.~\ref{multilayersetup}). The cavities
can be coplanar wave guides or microstrip resonators. Available multi-layer
technology allows the fabrication of thin films stacked on top of
each other. For example, consider waveguide layers of Nb or Al, separated
by $100\,{\rm nm}$ of dielectric, which can be optimized for good
decoherence properties (e.g.~\cite{2008_03_Martinis_Dielectrics}).
The inner grid area for $10\times10$ wave guides (each with $20\mu{\rm m}$
inner conductor and $10\mu{\rm m}$ gaps) would have a width of about
$1\,{\rm mm}$, whereas the full resonator lengths are above $10{\rm mm}$,
allowing all the qubits to be placed near the cavity mode central
field antinode, with comparable couplings (see Fig.~\ref{multilayersetup}).
The qubit-cavity coupling remains similar to single-layer designs,
owing to the small layer thickness of only $100\,{\rm nm}$. Good
isolation between two orthogonal cavities was estimated in a previous
theoretical work \cite{2006_12_Storcz_OrthogonalCavity,2008_WMI_QuantumSwitch}, and unwanted
cross-talk may be reduced further by choosing different cavity frequencies.
The qubits can be placed above all layers to minimize fabrication
problems. Weak coupling between qubits and control lines (e.g. cross-capacitance
$\sim0.01{\rm fF}$) suppresses sufficiently unwanted Nyquist noise from these
lines, which could lead to decoherence.  Indeed, 
for a Cooper pair box of total capacitance 
$C_{\Sigma}$
, this cross-capacitance yields a relaxation rate 
$\sim(C_{g}/C_{\Sigma})^{2}e^{2}\omega Z/\hbar$
 for radiation into a control line of impedance $Z$
 at the qubit splitting frequency 
$\omega$, leading to estimates that are small compared to the intrinsic qubit relaxation
rate for the present parameters (the same holds for dephasing).

\begin{figure}[t]
\includegraphics[width=\columnwidth]{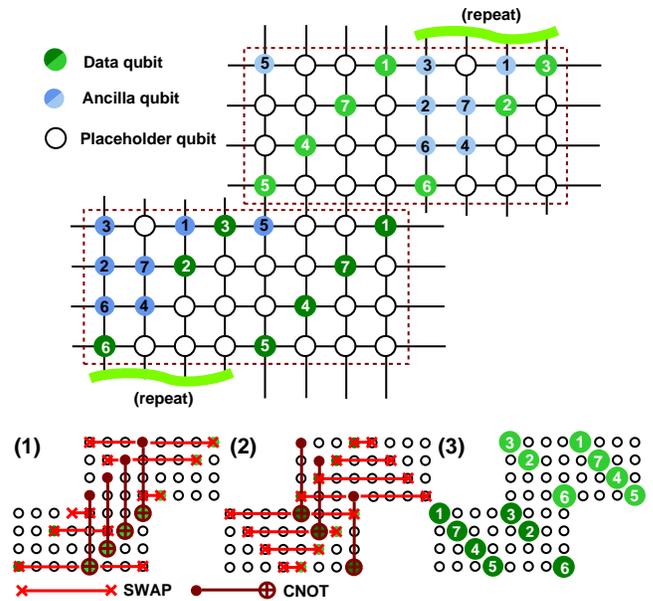}

\caption{\label{FaultTolerantFigure}A possible fault-tolerant scalable architecture
based on the cavity grid. Top: The unit cell of a periodic arrangement,
with two logical qubits, each made up of seven data qubits (grouping
indicated by dashed rectangles), together with ancilla and placeholder
qubits. Bottom: The sequence of SWAP and CNOT gates shown in (1) and
(2) implements a transversal CNOT between the logical qubits, producing
the final arrangement (3); see main text.}

\end{figure}

\section{Scalable, fault-tolerant architecture}
The cavity grid is
a building block for a truly scalable, fault-tolerant architecture.
Scalability means that, at a minimum, the physics of initialisation,
readout, single- and two-qubit gates does not depend on the total
number of qubits. Fig.~\ref{FaultTolerantFigure} shows a scalable
architecture requiring only eight different qubit frequencies. In
each unit cell of 64 qubits (Fig.~\ref{FaultTolerantFigure}) we
choose two arrays of seven data qubits and use each array to store
a single logical qubit, employing the Steane quantum error correction code \cite{1996_Steane_ErrorCorrection}. Clean
logical states can be prepared in additional ancilla qubits. Moreover,
errors in the data qubits can be copied into the ancillae, which are
then measured, locating the errors and enabling correction \cite{2007_DiVincenzo_QErrorCorrectionSlowMeasurement}.
All other qubits are placeholders, which are crucial: Swapping a pair
of data qubits directly could corrupt both if the SWAP gate fails,
resulting in a pair of errors that may not be correctable by the seven
qubit Steane code. Using three SWAP gates with a placeholder qubit
for temporary data storage solves this problem. We ignore errors in
placeholder qubits as they contain no data.

A logical CNOT gate is illustrated in Fig.~\ref{FaultTolerantFigure}.
The final arrangement of qubits differs from the initial one and can
be returned to it by swapping. However, if all logical qubits undergo
similar logical gates, explicitly swapping back may be unnecessary
as subsequent gates will do this automatically. A broad range of single
logical qubit gates are possible. Full details of our chosen set of
logical gates and their associated circuits including error correction
can be found elsewhere \cite{2007_02_Hollenberg_FaultTolerantQC,2007_04_FowlerWilhelm_ScalableFluxQubits}.

\section{Conclusions} In this Letter, we have proposed a novel architecture for quantum computation
using a 2D grid of superconducting qubits coupled to an array of on-chip
microwave cavities. A {}``Sudoku''-type arrangement of qubit frequencies
permits global coupling of a large number of qubits while suppressing
spurious interactions. These basic ideas could be implemented in a wide variety of  hardware implementations. Elementary operations within this scheme could
be demonstrated in the near future on small grids, while the setup
has the potential to form the basis for truly scalable fault-tolerant
architectures. 

\acknowledgments
We acknowledge useful discussions with R.~J.~Schoelkopf, A.~Imamo\u{g}lu,
A.~Blais, A.~Wallraff, H.~J.~Majer, F.~Deppe, Y.~Nakamura, D.~Esteve,
C.~Wilson and R.~Gross, and support by EuroSQIP and the DFG research
networks SFB 631 and NIM, and the Emmy-Noether program (F.M.). ES thanks 
the Ikerbasque Foundation, the EU EuroSQIP project, 
and the UPV-EHU grant GIU07/40.
\bibliographystyle{apsrev}
\bibliography{BibFM}

\end{document}